\documentclass[a4paper,fleqn]{cas-dc}

\usepackage[numbers]{natbib}
\usepackage{braket}
\usepackage{qcircuit}

\begin{document}
\let\WriteBookmarks\relax
\def\floatpagepagefraction{1}
\def\textpagefraction{.001}
\shorttitle{Adiabatic Quantum Algorithm for Multijet Clustering in High Energy Physics}
\shortauthors{D. Pires, Y. Omar and J. Seixas}

\title [mode = title]{Adiabatic Quantum Algorithm for Multijet Clustering in High Energy Physics}                      

\author[1]{Diogo Pires}
\ead{diogofgpires@tecnico.ulisboa.pt}

\author[1,2]{Yasser Omar}
\ead{yasser.omar@lx.it.pt}

\author[1,3]{João Seixas}
\ead{joao.seixas@tecnico.ulisboa.pt}

\address[1]{Instituto Superior T\'{e}cnico, Universidade de Lisboa, Portugal}
\address[2]{Instituto de Telecomunica\c{c}\~oes, Physics of Information and Quantum Technologies Group, Lisbon, Portugal}
\address[3]{Centro de Física e Engenharia de Materiais Avançados (CeFEMA), Instituto Superior Técnico, Av. Rovisco Pais 1, 1049-001 Lisboa, Portugal}

\begin{abstract}
The currently predicted increase in computational demand for the upcoming High-Luminosity Large Hadron Collider (HL-LHC) event reconstruction, and in particular jet clustering, is bound to challenge present day computing resources, becoming an even more complex combinatorial problem. In this paper, we show that quantum annealing can tackle dijet event clustering by introducing a novel quantum annealing binary clustering algorithm. The benchmarked efficiency is of the order of $96\%$, thus yielding substantial improvements over the current quantum state-of-the-art. Additionally, we also show how to generalize the proposed objective function into a more versatile form, capable of solving the clustering problem in multijet events.
\end{abstract}


\maketitle

\section{Introduction}

In high-energy physics jets play a fundamental role, signaling the presence of partons produced in the interaction, and providing us with valuable information regarding the underlying Quantum Chromodynamics (QCD) processes. When a quark-antiquark pair is produced, as the distance between quarks increases, the energy associated with this separation also increases. This means that for a sufficiently large distance the energy will eventually be large enough for a new, more energetically favorable quark-antiquark pair to be produced. Since quarks obey color-confinement the final hadronic states produced from these quarks must evolve to colorless bound states -- a process called hadronisation. These hadrons and subsequent final stable particles tend to travel all in the same direction, forming narrow, collimated sprays of particles - jets.

From the collection of final-state particles produced in a given event, jet clustering aims at finding which particles belong to which jet clusters by analysing these particles' properties in an approximate attempt to reverse-engineer the  underlying quantum mechanical QCD processes of fragmentation and hadronisation. A jet algorithm then maps the momenta of $N$ collimated and energetic final-state particles $\{\Vec{p}_i\}$, into the momenta of $K$ cluster jets $\{\Vec{j}_k\}$, dependent on the collision conditions and the particles' subsequent final-state geometry and distribution. 

Considering the HL-LHC upgrade currently ongoing, the computational resources demand is set to increase drastically, resulting in a predicted $\sim10$X increase in both pile-up, from $<\mu> \sim 20$ to $<\mu> \sim 200$, \cite{LHC, HLLHC} and subsequent produced particle multiplicity. As a consequence, event reconstruction, and in particular jet clustering, is bound to become an even more complex combinatorial problem, with a significant increase in final-state number of particles $N$ to be clustered. The amount of clustering possibilities will increase thus challenging present day computing resources. 

In this work, we study the possibility of using quantum annealing to tackle the problem of jet clustering, by introducing a new, angle-based quantum annealing formulation and establishing its performance. For this, we implement, for the first time, the quantum state-of-the-art by Wei \textit{et al.} \cite{Harrow} as well as the proposed quantum annealing algorithm, on the latest cloud-available \textit{D-Wave} annealing machine, the \textit{Advantage} Quantum Processing Unit (QPU) \cite{DWave}, hoping to understand how both algorithms perform and compare in terms of clustering efficiency. By benchmarking the obtained results against those of the widely used $k_t$ clustering algorithm \cite{kt}, we have shown that the proposed algorithm yields improved results over the quantum state-of-the-art \cite{Harrow}.

\section{Quantum Annealing}

Quantum annealing aims at solving global optimization problems, its primary goal being to find the minimum of a defined objective function. Making the analogy between the global minimum of this function and the ground state of a system, quantum annealing makes use of quantum tunneling processes to lead the system to its global minimum \cite{Annealing}.

At its core lies the adiabatic theorem \cite{Messiah}, which tells us that if the gap $E_1(t)-E_0(t)$ between the two lowest energy levels of our quantum system is strictly greater than zero for the entire annealing evolution time window and if the evolution is taken to be slow enough, our Schr{\"o}dinger equation obeying quantum state $\ket{\psi(t)}$ will then remain very close to the instantaneous ground state for all $t$ from $0\le t\le T$, where $T$ is the annealing time. We now define the minimum gap $g_{\min}$ as:

\begin{equation}
\begin{split}
    &g_{\min} = \min_{0 \leq s \leq 1} \big(E_1(s)-E_0(s)\big), \\ &\text{\small with} \quad s = t/T \quad  \text{\small and} \quad T \sim O(g_{\min}^{-2})~. 
\end{split}
    \label{eq: spectral_gap}
\end{equation}

Typically, the quantum system is initialized in the ground state of a simple and known hamiltonian $H_i$. We want to find out the ground state of another hamiltonian, $H_f$, which is rather simple to specify, but whose ground state turns out to be hard to find (this corresponds to the definition of our optimization problem/function to be minimized). We now perform the annealing slowly enough as to go from the known ground state of $H_i$ to the unknown ground state of $H_f$  \cite{Annealing}:
\begin{equation}
\begin{split}
    H(t) &= \left(1-\frac{t}{T}\right)H_i + \frac{t}{T}H_f \\ &\text{\small or} \\
    \widetilde{H}(s) &= (1-s)H_i + sH_f
\end{split}
    \label{eq: Anneal_Ham}
\end{equation}
By preparing the state in such way that at $t = 0$ it corresponds to the ground state of $H(0) = H_i$ and if $g_{\min} > 0$, then in the end of the annealing process, for large enough $T$, $\ket{\psi(t)}$ will be very close to the ground state of $H_f$, that is, to the solution of our optimization problem. 

There are two key formulations for the objective function regarding its input form for the \textit{D-Wave} computer: the \textit{Ising} and the Quadratic Unconstrained Binary Optimization (QUBO) formulations (with the possibility of trivial conversion through $s_i = 2x_i-1$). In the \textit{Ising} model, the variables $s_i$ take the values of either "spin up" ($\uparrow$) with $s_i = +1$, or "spin down" ($\downarrow$) with $s_i = -1$. The $N$-variable objective function expressed as an \textit{Ising} model takes the following form:
\begin{equation}
    E_{Ising}(s) = \sum_{i=1}^N h_i s_i + \sum_{i<j=1}^N J_{ij}s_is_j, 
    \quad s_i \in \{-1,+1\}~. 
    \label{eq: Ising_form}
\end{equation}
Equivalently, the QUBO formulation is defined by a $N \times N$ upper-triangular matrix, $Q$, of real weights, and a vector of binary variables $x$. The goal is then to minimize the function:
\begin{equation}
    E_{QUBO}(x) = \sum_i Q_{ii}x_i + \sum_{i<j} Q_{ij}x_ix_j \equiv \sum_{i,j=1}^N Q_{ij}x_ix_j~,
    \label{eq: QUBO_form}
\end{equation}
where $x_i \in \{0,1\}$. The second form of equation \eqref{eq: QUBO_form} is obtained by summing $i$ and $j$ with repeated indices given that $x^2_i = x_i$ allows us to absorb the linear terms into the quadratic terms.

%%%%%%%%%%%%%%%%%%%%%%%%%%%%%%%%%%%%%%%%%%%%%%%%%%%%%%%%%%%%%%%%

\section{Dijet Events}

To illustrate dijet events, we shall take the $e^+e^-$ collision case as an example. An electron and a positron are accelerated in opposite colliding directions on the same beam axis, up until the point at which they collide, annihilate (given that they possess opposite momenta, charge, and the same energy), and end up emitting a quark-antiquark pair, through either a  virtual $Z^0$ gauge boson or a virtual photon. 

This quark-antiquark pair then hadronizes, giving rise to highly collimated sprays of hadrons known as jets. Since the original state is composed of two highly energetic partons with opposite momenta, the $N$ produced hadrons will mostly give rise to $K=2$ jets. Taking as input this hadronic final-state, we now face the task of understanding which hadrons belong to which jet, grouping them together, and finally computing both jet's final momenta in order to gain insight into the original parton state.

We can take advantage of the geometrical properties of dijet events in order design an effective solution. It turns out that it is intuitive to formulate the task as an optimization problem, where we seek to minimize some cleverly defined quantity, leading directly to the desired particle-to-jet association. Once this appropriate quantity (here denoted by $Q$) that fits the underlying nature of the problem and which unveils the desired solution has been found, the next step is to write down the related objective function, or the function which is to be minimized/optimized. What does this quantity $Q$ depend on? What are the key variables and parameters that, once tweaked to the right values, will reveal the minimum value of $Q$ and hence guide us to the correct configuration of jets? This function then needs to be expressed in a certain way, typically in the form of equation \eqref{eq: Ising_form} or equation \eqref{eq: QUBO_form}, depending on both the nature of the problem at hand and on the defined quantity $Q$.

In the state-of-the-art work by Wei \textit{et al.} \cite{Harrow}, the \textit{Thrust} \cite{thrust} $T$ is used, taking advantage of the collimated nature of jets and claiming that the more "pencil-like" the momenta of the grouped particles are, the closer one is to the desired final jets configuration. By maximizing the \textit{Thrust} the authors aim at sorting the $N$ final-state particles to be clustered into two hemispheres, thus inherently tackling dijet events. Given the nature of its partitioning/hemisphere formulation, the authors further choose to express the problem in the QUBO form. In contrast, as we will see, we shall define and use a more general quantity.

%%%%%%%%%%%%%%%%%%%%%%%%%%%%%%%%%%%%%%%%%%%%%%%%%%%%%%%%%%%%%%%%

\section{Quantum Angle-Based Clustering}

We now introduce a novel quantum annealing formulation for jet clustering, aiming at describing the physical concept and rationale behind it. The results of the algorithm's implementation and the runs on the \textit{D-Wave Advantage}'s 5000-qubit QPU (available to the general public via cloud) \cite{DWave} are also presented and discussed according to its benchmarks, hoping to understand how well it performs relative to the state-of-the-art.

As opposed to the quantum state-of-the-art formulation \cite{Harrow}, the algorithm introduced here relies on a more general angle-based quantity, which can be applied to any kind of $K$-jet final-state in order to perform clustering. Unfortunately, given current hardware constraints, we aim at implementing the $e^+e^-$ collision event case, briefly introducing its $K$-jet generalization formulation in the end. 

\subsection{Algorithm}

We take a purely mathematical perspective, with the goal of mapping a collection of $N$ particles' momentum vectors $\{\Vec{p}_i\}$, corresponding to $N$ final-state particles, onto a set of output final jets, $\{\Vec{j}_k\}$ (here with $k \in \{1,2\}$). All these particles are assumed to originate from the same point in space, and should be sorted into the relevant jet clusters, adequately recombined into the jet's final total momenta, $\Vec{j}_k$. As such, starting from the assumption that $N$ particles are to be assigned to $K=2$ jets, it turns out to be  conceptually more intuitive to express the objective function in terms of \textit{Ising} variables $s_i = \pm 1$. This way, $s_i = 1$ indicates that a given particle $i$ belongs to jet cluster $j_1$, while $s_i = -1$ indicates that the same particle $i$ does not belong to $j_1$, thus belonging to the remaining jet cluster $j_2$. We start by writing a general objective function \textit{ansatz}: 
\begin{equation}
    H = \frac{1}{2} \sum_{i,j = 1}^N d(\Vec{p}_i,\Vec{p}_j) s_i s_j~, \label{eq: binary_Ising}
\end{equation} 
where $d(\Vec{p}_i,\Vec{p}_j)$ represents a dissimilarity metric, analogous to the quantity $Q$ mentioned above. Whenever the dissimilarity $d(\Vec{p}_i,\Vec{p}_j)$ between two particles $\Vec{p}_i$ and $\Vec{p}_j$ is large, $s_i$ and $s_j$ tend to take opposite signs, thus being assigned to different clusters. On the other hand, if $d(\Vec{p}_i,\Vec{p}_j)$ is small, $s_i$ and $s_j$ take the same value and are assigned to the same cluster. Since $s_i$ can never be set to zero there is only one $s_i$ per particle for the $N$ particles. This means that all particles are assigned to one and only one cluster. The factor $1/2$ accounts for the symmetric nature of the dissimilarity metric $d(\Vec{p}_i,\Vec{p}_j) = d(\Vec{p}_j,\Vec{p}_i)$ in the sum. Moreover, since we have an $s_i$ variable, and thus a qubit per particle, we end up with a qubit usage $O(N)$, representative of the $N$ final-state particles being clustered.

When choosing $d(\Vec{p}_i,\Vec{p}_j)$, it is important to note that a standard Euclidean distance metric would not be the best choice. This can be understood by picturing two soft particles with momenta $\Vec{p}_i$ and $\Vec{p}_j$ belonging to different jets. These are registered as being closer to the vertex than their hard companions, that is, they possess a smaller momentum norm relative to the others. In the cases where the energy gap is sufficiently large, the minimization process will be harmed since $d(\Vec{p}_i,\Vec{p}_j)$ is smaller relative to the average $\overline{d}$, thus erroneously grouping $\Vec{p}_i$ and $\Vec{p}_j$ together. 

We now search for a quantity which facilitates the minimization of its objective function, and is aligned with our views and goals for this problem: as this hypothetical $d(\Vec{p}_i,\Vec{p}_j)$ increases, the probability that the two particles belong to the same jet should be as correlated with it as possible. In other words, the larger/smaller the chosen quantity gets, the larger/smaller the output energy of the corresponding \textit{Ising} hamiltonian should be.

Given the high energy of the initial outgoing quark-antiquark pair, the final jets tend to be highly collimated, such that, in general, we have $\theta(\Vec{p}_i,\Vec{p}_j) \ll \frac{\pi}{2}$ for any two particles $\Vec{p}_i$ and $\Vec{p}_j$ in the same jet. One can thus leverage upon this important feature of jets by using the angle $\theta$ between particles as a starting point to build an appropriate dissimilarity metric. As such, for our hamiltonian, we write:
\begin{equation}
\begin{split}
    H &= \frac{1}{2} \sum_{i,j = 1}^N -\cos\big[\theta(\Vec{p}_i,\Vec{p}_j)\big]s_is_j \\
    &= \frac{1}{2} \sum_{i,j = 1}^N -\frac{\Vec{p}_i \cdot \Vec{p}_j}{|\Vec{p}_i| \cdot |\Vec{p}_j|} s_is_j ~,
    \label{eq: final_H} 
\end{split}
\end{equation} 
where $ d_{ij} = -\cos\big(\theta_{ij}\big)$. When particles $\Vec{p}_i$ and $\Vec{p}_j$ belong to the same jet, we measure $\theta_{ij} \ll \frac{\pi}{2}$, thus yielding $\cos\big(\theta_{ij}\big) \approx 1$. On the other hand, whenever two particles $\Vec{p}_i$ and $\Vec{p}_j$ belong to opposite jets), we measure $\theta_{ij} \sim \pi$, yielding $\cos(\theta_{ij}) \approx -1$. 

Because our goal is to minimize $H$ and not to maximize it, we introduce a minus sign in equation \eqref{eq: final_H}. As a result, the minimization of $H$ will therefore favor the clustering of particles closer in angular distance, that is, with smaller $\theta(\Vec{p}_i,\Vec{p}_j)$ relative to one another. This is exactly what we are looking for, as particles in the same jet tend to have significantly smaller angular separations when compared to particles in opposite jets. 

Even though equation \eqref{eq: final_H} refers to simpler cases of dijet events, the dissimilarity metric used is much more versatile and can be generalized to more complex events. As such, opposed to the \textit{Thrust} discussed in Wei \textit{et al.} \cite{Harrow}, we are therefore safe while carrying this concept to more elaborate, $K$-jet generalizations. 

\subsection{$K$-jet Generalization}

In order to generalize the above quantum annealing algorithm, one must first realize that the way the algorithm is formulated in equation \eqref{eq: final_H} does not allow for more than two jets ($K > 2$) per event. When $K > 2$, the two variables $s_i$ and $s_j$ cannot, on their own, sort a given particle to an hypothetical third jet: given that $s_i = \pm 1$, each of the two allowed states accounts for one jet each, meaning that each particle $\Vec{p}_i$ can only be sorted into either the jet corresponding to $s = 1$ or to the jet corresponding to $s = -1$.

When moving to a more elaborate $K$-jet event, it is intuitively clear that one needs $K$ binary variables for each particle in order to successfully sort any given particle to any of the $K$ jets. This can easily be done by assigning a positive variable value if the corresponding particle is to be sorted into the jet in question, and a negative (or zero) value for every other variable corresponding to every other jet. It is essential that each of the final-state particles is assigned to one and only one jet at a time, since particles are not physically allowed to be present in more than one jet.

Given the more complex nature of the problem at this point, we opt to change the way we express the objective function to be minimized. We shall use the usual binary variables $x^k_i$ and $x^k_j$ to denote whether or not two given particles $\Vec{p}_i$ and $\Vec{p}_j$ belong to the same jet $j_k$. As such, if particle $\Vec{p}_i$ is considered to be included in jet $j_k$, it will have $x^k_i = 1$. If not, it would have $x^k_i = 0$. To this type of formulation where we have one qubit per particle per jet, we call One-Hot Encoding \cite{kumar}. It comes at the cost of a more intensive qubit usage of the order of $O(KN)$. As such, we can start by writing the first term of our $K$-jet objective function:
\begin{equation}
   H'_K = \frac{1}{2} \sum_{k=1}^K \sum_{i,j=1}^N -\cos\big[\theta(\Vec{p}_i,\Vec{p}_j)\big] x_i^k x_j^k~. \label{eq: K_jet1}
\end{equation}
However, since now the lowest energy possible for a given configuration is zero, we know that the minimization process of the objective function favors the scenario in which all particles are assigned to zero jets, such that we have $x_i^k = 0$ either for a given particle $\Vec{p}_i$ and all jets $j_k$ with $k \in \{1,\dotsc, K\}$, or for a given jet $j_k$ and all particles $\Vec{p}_i$ with $i \in \{1,\dotsc, N\}$. Furthermore, as mentioned above, we must not allow for any given particle to be assigned to more than one jet. 

Both of these issues can be solved by adding an adequately built constraining term. Again, one needs to guarantee that for each particle $\Vec{p}_i$, there is one and only one $x_i^k = 1$ for some jet $j_k$, with the rest of $x_i^{k' \neq k} = 0$. We thus introduce
\begin{equation}
    \phi_i = \Bigg(1 - \sum_{k=1}^K x_i^k\Bigg)^2~, \label{eq: constraint}
\end{equation}
and add it with a tunable parameter $\lambda$ to \eqref{eq: K_jet1} in order to obtain the complete hamiltonian:
\begin{equation}
\begin{split}
    H_K :&=  H_K' + \lambda \sum_{i=1}^N \phi_i \\
    & = \frac{1}{2} \sum_{k=1}^K \sum_{i,j=1}^N -\cos\big[\theta(\Vec{p}_i,\Vec{p}_j)\big] x_i^k x_j^k  \\
    & + \lambda\sum_{i=1}^N\Bigg(1 - \sum_{k=1}^K x_i^k\Bigg)^2~. \label{eq: K_jet_final}
\end{split}
\end{equation} 

When it comes to defining the magnitude of $\lambda$, it is important to remember the reason that justifies the need for this constraining term just added to \eqref{eq: K_jet1}. As mentioned above, one needs to guarantee that for each particle with momentum $\Vec{p}_i$, one and only one $x_i^k = 1$ for some jet $j_k$ while the remaining $x_i^{k' \neq k} = 0$. If a particle is assigned to more than one jet, the constraining term grows with each additional jet the particle is assigned to. Consequently $H_K'$ will never energetically favor this possibility, since it will always result in an increase of its energy. However, in the remaining case in which a given number of particles are assigned to zero jets, the corresponding first terms of the hamiltonian will be set to zero and reduce the value of $H_K$, thus being energetically favored. One can thus say that the goal to be achieved with the addition of ´the constraint \eqref{eq: constraint}, is simply to offset the largest possible "incorrect" energy reduction in $H_K'$. When a particle $\Vec{p}_i$ is assigned to zero jets, it can, in a worst case scenario basis, result in $N-K$ pairwise dissimilarity metrics set to zero \cite{kumar}. Furthermore, the maximum possible reduction in $H_K'$ would correspond to when each of the remaining particles are at a maximum (angular) distance from particle $\Vec{p}_i$. In such a setting, we can write:
\begin{equation}
    \sum_{\Vec{p}_j \in j_k} d(\Vec{p}_i,\Vec{p}_j) \leq (N-K) \cdot \max_{\Vec{p}_j \in j_k} d(\Vec{p}_i,\Vec{p}_j)~. \label{eq: lambda_size}
\end{equation}

We are now in conditions to conclude that the approximate order of magnitude for $\lambda$ should be
\begin{equation}
    \lambda \sim (N-K) \cdot \max \Big( -\cos \big[\theta(\Vec{p}_i,\Vec{p}_j)\big]\Big), \quad \forall \Vec{p}_i,\Vec{p}_j~. \label{lambda_size_final}
\end{equation}
It is important to mention that in practice, and even though sometimes desirable, $\lambda$ cannot be made arbitrarily large due to the current hardware state of the art inherent limitations mainly related to the allowed range of the qubits' couplings \cite{Harrow}. As such, it should be pointed out that when compared to the $K=2$ jet event, the $K$-jet one-hot encoding formulation is considerably harder to implement on current quantum annealing hardware, with previous numerical studies \cite{kumar} having shown that clustering problems making use of multiple qubits to implement one-hot encoding are prone to errors, thus widening the performance gap between dijet and multijet events.

%%%%%%%%%%%%%%%%%%%%%%%%%%%%%%%%%%%%%%%%%%%%%%%%%%%%%%%%%%%%%%%%

\section{Implementation}

\subsection{QPU Inputs}

In the standard QUBO formulation, the QPU takes as input the matrix elements $Q_{ij}$, as seen in equation \eqref{eq: QUBO_form}. However, in the case of an \textit{Ising} formulation, the input values are $h_i$ and $J_{ij}$, according to equation \eqref{eq: Ising_form}.

% In order to successfully connect to the QPU and implement equation \eqref{eq: final_H}, one needs to express it in terms of a few essential terms. In the case of a QUBO formulation of the problem, this would correspond to the $Q$ matrix mentioned above. In the case of our algorithm, however, we have expressed it in the \textit{Ising} form, such that the terms now required are both the vector $h$, but also the coupling matrix $J$.

Since the proposed algorithm has been formulated in the \textit{Ising} form, by looking at equation \eqref{eq: final_H} and comparing it to equation \eqref{eq: Ising_form}, we see that there is no term of the form $s_i$, but only a term of form $s_is_j$. As such, we have the $s_i$ corresponding factor $h_i = 0$. Similarly, we immediately obtain the following expression for $J_{ij}$:
\begin{equation}
    J_{ij} = -\frac{1}{2} \frac{\Vec{p}_i \cdot \Vec{p}_j}{|\Vec{p}_i| \cdot |\Vec{p}_j|}~. \label{eq: J_matrix}
\end{equation}

After conducting a series of small runs to determine the performance's sensitivity to the annealing parameters, we have chosen to run the quantum annealing algorithm with the default \textit{annealing\_time} parameter set to $\text{\textit{annealing\_time}} = 20 \mu s$. Furthermore, we have also set the \textit{num\_reads} parameter, which defines the number of anneals performed, to $\text{\textit{num\_reads}} = 5000$, in order to have a reasonable amount of accumulated statistics resulting in a good balance between running time and accuracy.

\subsection{Event Generation}

The \textit{PYTHIA} Monte-Carlo event generator \cite{pythia} (version 8.3) was used to as realistically as possible simulate real data. Given the $K=2$ binary nature of the jet events being studied, we generated $e^+e^- \rightarrow Z^0 \rightarrow q\bar{q}$, with all $Z^0$ decays switched off with only those to quarks having been manually switched on. Using the data output by \textit{PYTHIA}, we then process it using a program specially designed to calculate $J$ according to equation \eqref{eq: J_matrix}.

%  Implicit by the center-of-mass energy used of $\sqrt{s} = m_Z$, only $q\bar{q} \in \{u\bar{u},d\bar{d},c\bar{c},s\bar{s},b\bar{b}\}$ decays have been allowed, since the $t$ quark is too massive for the center-of-mass energy used here ($m_t \ll \sqrt{s} = m_Z$).

\subsection{Benchmarks}

When it comes to measuring the quantum algorithms' jet clustering quality, it would be ideal to compare them to some implicit jet regrouping rules for any given generated event. Unfortunately such Monte-Carlo generated information is not available by design. Consequently, we have chosen to measure the algorithms' performance against that of the classical state-of-the-art $k_t$ clustering algorithm \cite{kt}.

The $k_t$ clustering algorithm has been implemented and used through the \textit{FastJet} software package \cite{Fastjet,kt_scaling}. By using the Jet Definition \textit{jet\_def(kt\_algorithm, R)}, the $k_t$ clustering algorithm \cite{kt} has been selected and chosen to run with an $R$ parameter of $R = 0.8$. Its output, was then a list of the final jets' total transverse momenta $||\Vec{j}_{T_k}||$, its pseudorapidity $\eta_k$ and the corresponding azimuthal angle $\phi_k$. In addition, the list of the regrouped final-state particles for each final jet was also produced, so that it could be used to compare the classical benchmark's results with those of the developed quantum algorithm.

\subsubsection{Clustering Efficiency}

For the purpose of measuring the algorithm's clustering quality, we have created an efficiency metric, $\epsilon$, adequately developed to serve our purpose. It is important to realize that the \textit{PYTHIA} generated $e^+e^-$ events are not bound to $K=2$ events despite being the most common. Indeed, one could observe, even though with smaller probability, $K=3$ or even $K=4$ events (\textit{e.g.} due to gluon radiation) within the generated data sets. As such, given that the developed algorithm is meant to be applied to binary clustering dijet events where $K=2$, we have made the choice of always considering only the two highest $p_T$ jets obtained by the $k_t$ benchmark for comparison with the (always) binary results obtained by the quantum annealing algorithms. Consequently, we developed the following efficiency metric to evaluate the obtained results for a given event $n$:
\begin{equation}
    \epsilon^{(n)} = \frac{\text{\# of particles grouped in the same way as $k_t$}}{\text{\# of particles in the two highest-$p_T$ jets ($k_t$)}}~. \label{eq: eff_metric}
\end{equation}
We have thus obtained two different total clustering efficiencies for $n$ generated events, $\epsilon_{\text{QBC}}$ and $\epsilon_{\text{Thr}}$, which reflect the efficiencies of the proposed quantum binary clustering algorithm and of the Thrust-based quantum annealing of Wei \textit{et al.} \cite{Harrow}, respectively:
\begin{equation}
    \epsilon_{\text{QBC}} = \frac{1}{n}\sum_n \varepsilon^{(n)}_{\text{QBC}}~, \qquad \epsilon_{\text{Thr}} = \frac{1}{n}\sum_n \varepsilon^{(n)}_{\text{Thr}}~.
    \label{eq: eff_total}
\end{equation}

\section{Results}

\begin{figure}[ht]
    \centering
    \includegraphics[scale = 0.40]{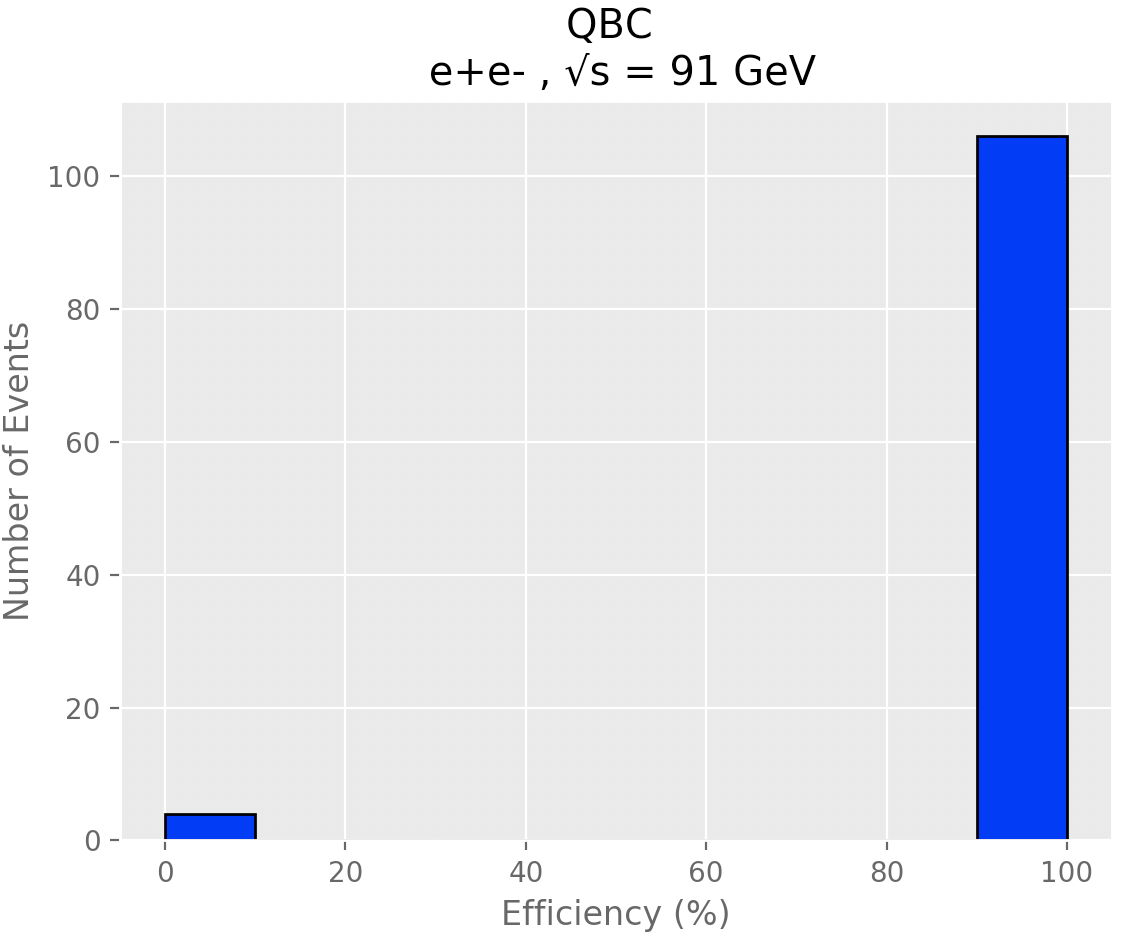}
    \caption{Histogram of the obtained efficiency for the proposed quantum binary clustering algorithm, $\epsilon_{\text{QBC}}$.}
    \label{fig:performance1}
\end{figure}

\begin{figure}[ht]
    \centering
    \includegraphics[scale = 0.40]{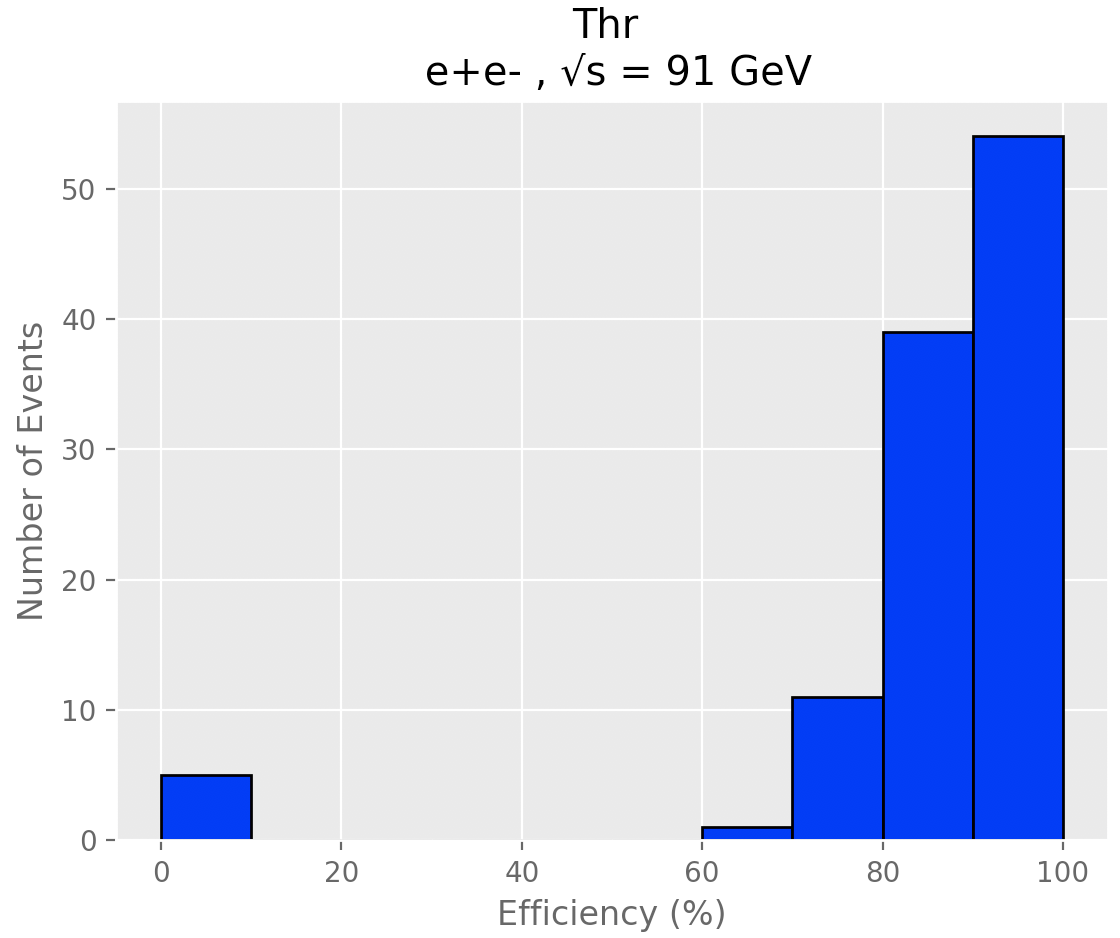}
    \caption{Histogram of the obtained efficiency for the Thrust-based algorithm by Wei \textit{et al.} \cite{Harrow}, $\epsilon_{\text{Thr}}$.}
    \label{fig:performance2}
\end{figure}

Given the limited amount of QPU time allowed for use on the \textit{D-Wave} machine, we have only been able to run the algorithms on a maximum number of $n = 110$ \textit{PYTHIA} generated $e^+e^-$ collision events. The resulting efficiency plots can be seen in Figures \ref{fig:performance1} and \ref{fig:performance2}.

Taking into account the small number of events generated, we have obtained for the proposed quantum binary clustering annealing algorithm, an efficiency of  $\epsilon_{\text{QBC}} = 96\%$. This efficiency $\epsilon_{\text{QBC}}$ has been obtained by computing each event-level efficiency through equation \eqref{eq: eff_metric}, and then finding the mean of all those efficiencies through equation \eqref{eq: eff_total}. In a similar procedure, for the \textit{Thrust}-based quantum annealing algorithm \cite{Harrow}, we have obtained an efficiency of $\epsilon_{\text{Thr}} = 85\%$. Moreover, it can also be seen from Figures \ref{fig:performance1} and \ref{fig:performance2} that the obtained efficiencies per event are much more stable for the quantum algorithm proposed, since almost $100\%$ of clustered events yield $\epsilon^{(n)}_{\text{QBC}} = 100\%$, whereas for the state-of-the-art quantum algorithm \cite{Harrow} that percentage is much smaller, closer to $50\%$.

Therefore, we can conclude that the proposed quantum binary clustering algorithm formulation yields superior results, resulting in a greater number of particles clustered in the same way as the $k_t$ algorithm, according to the efficiency metric developed of equation \eqref{eq: eff_metric}. This improvement can be explained not only by the inherently different clustering metric introduced, based on the angular separation of the particles being clustered, but also by the resulting \textit{Ising} formulation, which differs from the original QUBO being used, and tends to have significant impact due to its hardware implementation.

%TC:ignore

\section{Conclusions}

By focusing on an angular distance-based approach for jet clustering, we have introduced a new quantum annealing algorithm that performs binary clustering, thus being designed especially for the dijet event case. Nonetheless, it has been formulated having in mind the possibility of its use for a more generic $K$-jet event. This generalization has been shown to be reachable, although not yet implementable due to the current state of the art hardware constraints.

Upon implementing and running the proposed algorithm on the \textit{D-Wave}'s QPU, we compared its performance to that of the Thrust-based quantum annealing algorithm proposed in Wei \textit{et al.} \cite{Harrow}, successfully demonstrating that it yields improved results ($\epsilon_{\text{QBC}} = 96\%$ versus $\epsilon_{\text{Thr}} = 85\%$) according to the developed efficiency metric, resulting in a greater total number of particles clustered in the same way as the $k_t$ benchmark.

As such, making use of a more intuitive angle-based metric, and despite being modestly introduced in a $K=2$ context, the introduced algorithm has shown superior results relative to the available quantum state-of-the-art. Furthermore, it has also been shown to be easily generalizable into the more complex case of the $K$-jet event, thus proving once again that quantum annealing is a suitable choice for future use in highly particle-dense HEP environments such as the projected HL-LHC. 

On a final note, it is important to notice that there is currently no viable option for jet clustering when it comes to digital quantum computation. Although here focusing on quantum annealing techniques, we strongly recognize the need for new digital quantum computation solutions for jet clustering. This would bring an entirely new approach to the task at hand, allowing us to better assess performance and scaling in comparison to the classical state-of-the-art, and further reinforcing the idea that quantum computation is indeed a suitable option for the high-energy physics realm.

\bibliographystyle{elsarticle-num}
\bibliography{cas-dc-template}

\section{Acknowledgements}
The authors would like to thank Duarte Magano for all the valuable feedback provided throughout the writing of this paper. YO thanks the support from Funda\c{c}\~{a}o para a Ci\^{e}ncia e a Tecnologia (Portugal), namely through project UIDB/50008/2020, as well as from projects TheBlinQC and QuantHEP supported by the EU H2020 QuantERA ERA-NET Cofund in Quantum Technologies and by FCT (QuantERA/0001/2017 and QuantERA/0001/2019, respectively), and from the EU H2020 Quantum Flagship project QMiCS (820505). JS would like to thank the support of FCT under contracts CERN/FIS-COM/0036/2019 and UIDB/04540/\linebreak 2020.

% \section{Author contributions statement}

% Must include all authors, identified by initials, for example:
% A.A. conceived the experiment(s),  A.A. and B.A. conducted the experiment(s), C.A. and D.A. analysed the results.  All authors reviewed the manuscript. 

% \section{Competing Interests}

% The corresponding author is responsible for submitting a \href{http://www.nature.com/srep/policies/index.html#competing}{competing interests statement} on behalf of all authors of the paper. This statement must be included in the submitted article file.

% \section{Additional information}

%TC:endignore

\end{document}